\documentclass[amssymb,prl,aps,twocolumn,amsmath,showpacs,superscriptaddress]{revtex4}

\usepackage[dvips]{graphicx}

\begin{document}

\title{Supercurrent-induced temperature gradient across a nonequilibrium SNS Josephson junction}
\author{M.S. Crosser}
\affiliation{Department of Physics and Astronomy, Michigan State
University, East Lansing, MI 48824-2320, USA}

\author{Pauli Virtanen}

\author{Tero T. Heikkil\"{a}}

\affiliation{Low Temperature Laboratory, Helsinki University of
Technology, P. O. Box 2200, FIN-02015 TKK, Finland}

\author{Norman O. Birge}

\email{birge@pa.msu.edu}

\affiliation{Department of Physics and Astronomy, Michigan State
University, East Lansing, MI 48824-2320, USA}

\date{\today}

\begin{abstract}
Using tunneling spectroscopy, we have measured the local electron
energy distribution function in the normal part of a
superconductor-normal metal-superconductor (SNS) Josephson
junction containing an extra lead to a normal reservoir.  In the
presence of simultaneous supercurrent and injected quasiparticle
current, the distribution function exhibits a sharp feature at
very low energy.  The feature is odd in energy, and odd under
reversal of either the supercurrent or the quasiparticle current
direction.  The feature represents an effective temperature
gradient across the SNS Josephson junction that is controllable by
the supercurrent.
\end{abstract}

 \maketitle

The study of nonequilibrium phenomena in superconductors has a
rich history extending over several decades
\cite{SuperconBackground}. Advances in microfabrication techniques
in the 1990's have broadened the range of possible experiments. In
1999, Baselmans \textit{et al.} \cite{BaselmansNature} turned a
superconducting-normal-superconducting Josephson junction (SNS JJ)
into a ``$\pi$-junction" by driving the electron energy
distribution far from equilibrium with a normal control current
applied perpendicular to the supercurrent in a four-terminal
geometry. Later, Huang \textit{et al.} \cite{Huang:02} produced
similar behavior in a three-terminal sample with the control
current injected from a single normal reservoir. The physical
explanation behind the $\pi$-junction
\cite{Yip:98}\cite{Wilhelm:98} is the same in the two geometries:
the electron energy distribution in the presence of the control
current has a double-step shape \cite{Pothier:97}, which
selectively populates the Andreev bound states that carry current
in the opposite direction to that in equilibrium.

While the physics of the three-terminal and four-terminal
$\pi$-junctions is similar, the current flow and energy
distributions are different in the two geometries. In the
four-terminal experiment, the control current and supercurrent
coexist only in the region where the two wires cross. In the
three-terminal experiment, the wire joining the two
superconductors carries both supercurrent and normal
(quasiparticle) current \cite{Shaikhaidarov:00}.  It was predicted
in Ref.~\onlinecite{Tero:03} that this coexistence of
quasiparticle current and supercurrent will lead to a left-right
asymmetry in the effective temperature, due to mixing of the even
and odd components of the distribution function, $f(E)$. In this
process, supercurrent redistributes the Joule heat due to the
quasiparticle current between the different parts of the system.
As a result, the double-step distribution function is distorted,
and in the absence of inelastic scattering, this distortion is
approximately proportional to the energy spectrum of the
supercurrent-carrying states. We present here the first
measurement of the distortion in $f(E)$, using a local tunnel
probe.

\begin{figure}[ptbh]
\begin{center}
\includegraphics[width=3.2in]{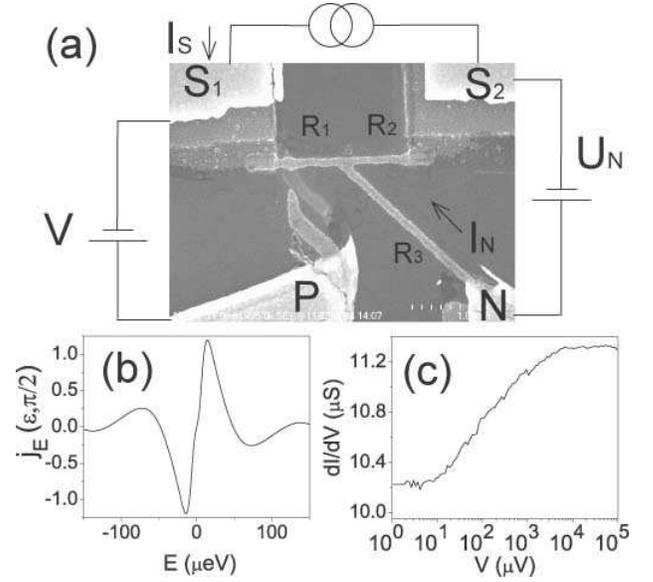}
\end{center}
\caption{(a) SEM micrograph of a sample and schematic measurement
circuit diagram. (b) Spectral supercurrent $j_E$ for $\phi = \pi
/2$ calculated numerically for the sample shown in (a).  (c) dI/dV
of the tunnel probe (P) measured at B=0.4 T, where
superconductivity is suppressed.  The 10\% reduction in dI/dV at
low bias is due to dynamical Coulomb blockade, as discussed in the
text.} \label{Fig1}
\end{figure}

Fig.~1a shows the three-terminal sample geometry. A normal metal
wire connects two superconducting electrodes ($S_1$ and $S_2$) and
a normal reservoir ($N$). The superconducting electrodes are kept
at the same chemical potential, which we define as the zero of
energy. With the normal electrode biased at a potential $U_N$, the
distribution function $f(E)$ can be calculated easily in the
absence of supercurrent, energy relaxation, and proximity effects.
In the normal reservoir, $f(E) = f_{FD}(E+eU_N)$ is a Fermi-Dirac
function displaced in energy by $-eU_N$.  Define the even and odd
(in energy) components of $f(E)$ as follows: $f_T(E) \equiv
1-f(E)-f(-E)$ and $f_L \equiv f(-E) - f(E)$.  The boundary
conditions at the NS interfaces for energies below the
superconducting gap, $\Delta$, are $f_T = 0$ and ${\partial
f_L}/{\partial x} = 0$, assuming high-transparency interfaces, no
charge imbalance in the superconductors, and no heat transport
into the superconductors \cite{andreev:64}. The solution for
$f(E)$ at the NS interface is the double-step shape: $f(E) =
0.5*[f_{FD}(E+eU_N) + f_{FD}(E-eU_N)]$.

In the presence of proximity effects and supercurrent, the kinetic
equations become more complicated \cite{Virtanen:04}:
\begin{subequations}
\begin{equation}
\frac{\partial j_T}{\partial x} = 0, \; j_T \equiv D_T(x)
\frac{\partial f_T}{\partial x} + j_E f_L +  \emph{T} (x)
\frac{\partial f_L}{\partial x} ;
\end{equation}
\begin{equation}
\frac{\partial j_L}{\partial x} = 0, \;\; j_L \equiv D_L(x)
\frac{\partial f_L}{\partial x} + j_Ef_T - \emph{T}(x)
\frac{\partial f_T}{\partial x} ;
\end{equation}
\label{eq:kinetic}
\end{subequations}
The spectral charge and energy currents, $j_T(E)$ and $j_L(E)$,
are conserved only in the absence of energy relaxation.  The
energy-dependent coefficients $D_T$, $D_L$, $j_E$, and $T$ can be
calculated from the Usadel equation \cite{usadel:70}, and vary
with the superconducting phase difference $\phi$ between $S_1$ and
$S_2$.  Fig.~1b shows the spectral supercurrent density, $j_E$,
for $\phi = \pi /2$ obtained from a numerical solution of the
Usadel equation appropriate for our sample.  In the presence of
the proximity effect, the kinetic equations must be solved
numerically. But one can understand the origin of the effective
temperature gradient predicted in Ref.~\onlinecite{Tero:03} by
noting that the terms with $j_E$ mix $f_T$ and $f_L$ in the
kinetic equations. If we neglect the $\emph{T}$ terms and ignore
the energy dependence in $D_T$ and $D_L$, then Eqs.~(1a) and (1b)
are easily solved analytically. The solution shows that $f_L$
along the horizontal wire connecting the two superconductors
contains a spatially antisymmetric contribution proportional to
$j_E$.

The experiments described in this paper were performed on four
samples, with similar results obtained on all four. Most of the
data reported here were taken from the sample shown in Fig.~1a,
with $R_1 = R_2 = 7.0~\Omega$, $R_3 = 16.9~\Omega$ and tunnel
resistance $R_T = 115~k\Omega$. The samples were fabricated using
a PMMA stencil patterned by electron-beam lithography and
multiple-angle evaporations \cite{Dolan:1988}. The three-legged
normal Ag wire is $70$~nm wide and $30$~nm thick. The
superconducting Al electrodes are $60$~nm thick, separated by a
distance of $1.4~\mu$m. The normal reservoir is $200$~nm thick.
The tunnel probe (connected to the pad labelled P) is
$20$~nm-thick Al oxidized in $80$~mbar of Ar(90\%)-O$_2$(10\%) for
6 minutes before deposition of the Ag wire.

We use tunneling spectroscopy to measure the shape of the
distribution function of the Ag wire at a position close to the
superconducting reservoir $S_1$, in a manner similar to that in
Ref.~\onlinecite{Pothier:97}.  A complication in our experiment is
that the density of states in the Ag wire is strongly modified by
proximity to $S_1$ and $S_2$. The current-voltage characteristic
of the probe tunnel junction is
\begin{eqnarray} \label{CBconduct}
I(V) =&-&\frac{1}{eR_T}\int dE n_S(E)\int d\varepsilon P(\varepsilon)\\
& &[f_S(E)n_N(E-eV-\varepsilon)(1-f_N(E-eV-\varepsilon))\nonumber\\
& -& (1-f_S(E))
n_N(E-eV+\varepsilon)f_N(E-eV+\varepsilon)]\nonumber
\end{eqnarray}
where $R_T$ is the normal state tunnel resistance, $n_{N}$ and
$n_{S}$ are the normalized densities of states, and $f_{N}$ and
$f_{S}$ are the electron energy distribution functions on the Ag
and Al sides of the tunnel junction, respectively. Eq.\
(\ref{CBconduct}) includes the effects of ``dynamical Coulomb
blockade" \cite{IngoldNazarov} through the quantity
$P(\varepsilon)$, the probability for an electron to lose energy
$\varepsilon$ to the resistive environment while tunneling through
the barrier.

Our strategy was to perform a series of experiments in which all
of the quantities in Eq.\ (\ref{CBconduct}) are known except for
one, and then deconvolve the data to find the missing quantity.
The first step is to find $P(\varepsilon)$, which can be
determined from $dI/dV(V)$ data with a large magnetic field $B$ to
suppress superconductivity.  Fig.~1c shows $dI/dV$ vs. $V$ with
$B$ = 0.4 T. In agreement with previous work \cite{anthore:03},
$dI/dV$ depends logarithmically on $V$ over nearly three decades
of voltage. For a tunnel junction in a resistive environment $R$,
with $R\ll R_K = h/e^2$, the function $P(\varepsilon)$ is well
approximated by:
\begin{equation}
P(\varepsilon)= \left\{
\begin{array}{cc}
 \frac{\alpha}{E_0} \left( \frac{\varepsilon}{E_0} \right)^{\alpha-1} &, \varepsilon<E_0 \\
 0 &,  \varepsilon>E_0
 \end{array} \right.\
\end{equation}
where $\alpha = 2R/R_K$ and $E_0 = e^2/(\pi \alpha C)$. The dI/dV
data in Fig.~1c can be fit to this form for $P(\varepsilon)$ with
$\alpha = 0.0176$ and $E_0 = 2.0$~meV, which correspond to the
reasonable values $R$ = 230 $\Omega$ and $C$ = 1.4 fF.

\begin{figure}[ptbh]
\begin{center}
\includegraphics[width=3.2in]{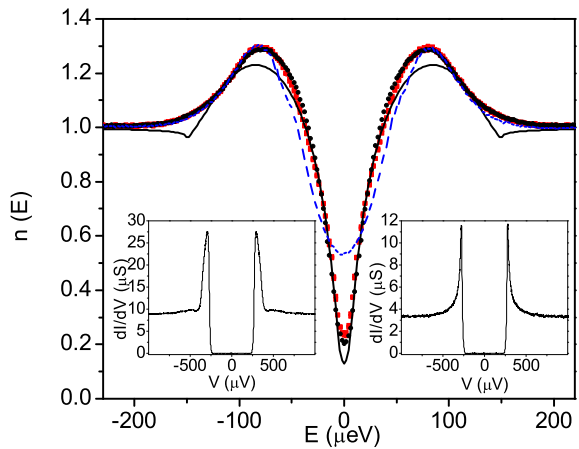}
\end{center}
\caption{(Color online) The $n_N(E)$ of the Ag wire at the
position of the probe electrode, with (red) and without (black)
supercurrent passing through the SNS Josephson junction, obtained
from dI/dV data such as shown in the left inset.  The solid line
is a fit to the solution of the Usadel equation discussed in the
text. The (blue) dashed line shows $n_N(E)$ needed to fit
nonequilibrium data with $U_N=53~\mu$V. Right inset: dI/dV data on
a reference SIN tunnel junction, used to determine the gap
$\Delta$ of the superconducting probe.} \label{Fig2}
\end{figure}

The second step is to determine $n_N(E)$, the density of states of
the Ag wire in close proximity to $S_1$.  In equilibrium, $f_S(E)$
and $f_N(E)$ are Fermi-Dirac functions at a temperature close to
that of the cryostat ($35$~mK). In most of our samples, the
$dI/dV(V)$ data in zero applied magnetic field show an anomalous
sharp feature at $eV \approx \Delta$, which disappears when
$B>10$~mT \cite{Fredthesis}; hence, all the low-field data
reported in this paper were taken with $B=12.5$~mT. We use the
standard BCS form for $n_S(E)$ of the Al probe, with a small
depairing parameter to account for the applied magnetic field
\cite{anthore:03b}. To determine $n_N(E)$ without ambiguity, we
must know the gap $\Delta$ of the Al probe. For that purpose we
fabricated, on the same chip, a reference SIN tunnel junction with
no proximity effect on the N side. The $dI/dV$ data for the SIN
junction, shown in the right inset of Fig.~\ref{Fig2}, could be
fit perfectly with Eq.\ (\ref{CBconduct}) assuming a BCS form for
$n_S(E)$ with $\Delta = 274~\mu$eV, a depairing parameter of
$\gamma \equiv \Gamma/\Delta =0.0020$ and with Dynamical Coulomb
Blockade parameters determined from high-field measurements on
that junction.  The rather large value of $\Delta$ is common for
thin thermally-evaporated Al films \cite{dirtyAl}. $dI/dV$ data
for the sample tunnel junction, shown in the left inset to Fig.~2,
looks noticeably different from the $dI/dV$ data on the reference
SIN tunnel junction due to proximity effect in the Ag wire. With
$\Delta$ determined, the $dI/dV$ data can be deconvolved to
produce $n_N(E)$ of the Ag wire, shown in the main panel of Fig.\
2. Equilibrium $dI/dV$ measurements were repeated with
supercurrent passing through the SNS JJ, since $n_N(E)$ depends on
$\phi$, the phase difference between $S_1$ and $S_2$. Fig.\ 2 also
shows $n_N(E)$ obtained with $I_S = 0.9 I_c$ and $-0.9 I_c$, which
are indistinguishable from each other.

To obtain a decent fit of the data in Fig.\ \ref{Fig2} to the
density of states calculated from the Usadel theory, we need to
assume a considerably smaller value of $\Delta\approx150~\mu$eV in
the superconducting reservoirs feeding the supercurrent than that
measured with the SIN reference junction. The reason for this
smaller value is the presence of a normal-metal--superconductor
bilayer close to these reservoirs (see Fig.\ \ref{Fig1}). Once we
know this $\Delta$, we find the Thouless energy $E_T=\hbar
D/L_S^2$ characterizing the normal-metal wire of length $L_S$
between the superconductors by fitting the temperature dependent
supercurrent at $U_N=0$ analogously to Ref.~\onlinecite{Huang:02}.
The quality of the fit is fairly insensitive to the precise value
of $\Delta$, and we obtain $E_T=5.56~\mu$eV.  In the fit, we also
corrected for the finite size of the superconducting terminals and
calculated the position-dependent order parameter $\Delta$
self-consistently close to the NS interface. Once we know these
two energy scales, we can calculate the local density of states
$n_N(E)$ at the position of the probe. As there is some ambiguity
about the position due to the finite normal-metal--superconductor
overlap, we fit also this within the limits set by the sample
geometry. This fit is shown as a solid line in Fig.~\ref{Fig2}.

\begin{figure}[ptbh]
\begin{center}
\includegraphics[width=3.2in]{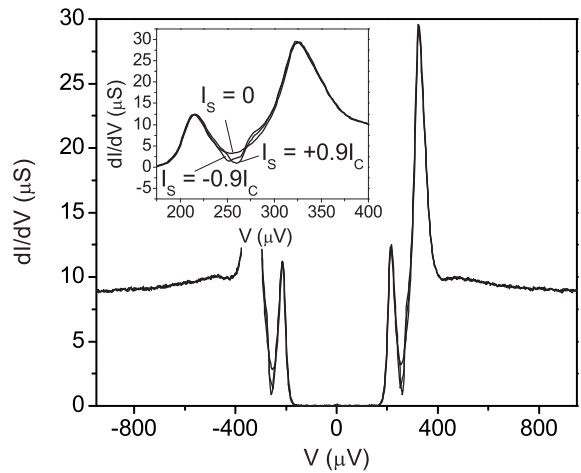}
\end{center}
\caption{dI/dV data of the probe tunnel junction with the normal
reservoir biased at $U_N = 53~\mu$V.  Inset: Enlarged view of the
double-peak at positive bias, showing the differences between the
data sets taken with $I_S = 0$ and $\pm 0.9 I_c$.} \label{Fig3}
\end{figure}

The third step in the experiment is to apply a voltage $U_N$ to
the normal reservoir, and repeat the dI/dV measurements across the
probe junction with various values of $U_N$ and $I_S$.
Fig.~\ref{Fig3} shows three sets of $dI/dV$ vs. $V$ data with $U_N
= 53~\mu$V, and with $I_S = 0, -0.9I_c$, and $+0.9I_c$.  The
double peaks in $dI/dV$ near $V = \pm \Delta$ arise from the
double-step shape in $f_N(E)$, the distribution function in the Ag
wire. Since we know $P(\varepsilon)$, $n_S(E)$, and $n_N(E)$, and
since $f_S(E)$ is a Fermi-Dirac function, we planned to determine
$f_N(E)$ from a deconvolution of these $dI/dV$ data. Instead, we
found that we were unable to fit the data shown in Fig.~\ref{Fig3}
using Eq.~(\ref{CBconduct}), with any physically permissible form
for $f_N(E)$. Only by allowing $n_N(E)$ to vary slightly from the
form obtained in equilibrium (dashed line in Fig. 2), could we fit
the nonequilibrium data.

\begin{figure}[ptbh]
\begin{center}
\includegraphics[width=3.4in]{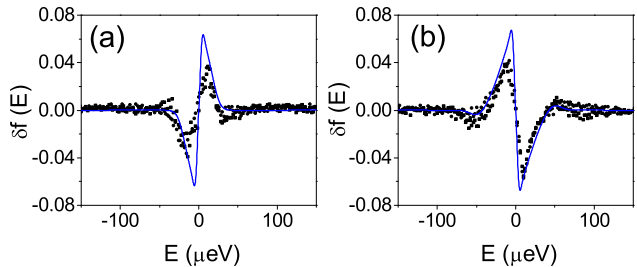}
\end{center}
\caption{(a) $\delta f(E) \equiv f_N(E,U_N,I_S) - f_N(E,U_N,-I_S)$
for $U_N = 22~\mu V$ and $I_S = 0.9 I_c$. (b) Same quantity for
$U_N = 53~\mu$V and $I_S = 0.9 I_c$, with the SNS JJ in the $\pi$
state. Notice the sign change between the two data sets. In both
figures, a second data set is shown with the signs of both $U_N$
and $I_S$ reversed. Solid lines are numerical solutions to
Eq.~(1).} \label{Fig4}
\end{figure}

Fortunately, we can extract the supercurrent-dependent changes in
$f_N(E)$ by considering differences between data sets with
opposite values of $I_S$. The equilibrium measurements shown in
Fig.~2 confirm our expectation that $n_N(E,U_N=0,I_S)$ =
$n_N(E,U_N=0,-I_S)$. This symmetry holds approximately also for
$U_N \neq 0$; hence the quantity $dI/dV(V,U_N,I_S) -
dI/dV(V,U_N,-I_S)$ depends only on the difference $\delta f_N(E)
\equiv f_N(E,U_N,I_S) - f_N(E,U_N,-I_S)=\delta f_L(E)$, the change
in the function $f_L(E)$. The left panel of Fig.~\ref{Fig4} shows
$\delta f_N$ defined in this way, with $U_N = 22~\mu$V and $I_S =
0.9 I_c$. There are sharp peaks at $E \approx \pm 15 \mu$V, odd in
energy.  This feature in $f_N(E)$ represents changes in the local
effective electron temperature, according to Eq. (5) in Ref.
\cite{Tero:03}. To confirm that the peaks are not due to a
geometrical asymmetry in the sample, such as a flaw in the
fabrication process, a second determination of $\delta f_N(E)$ is
shown superposed on the first. This second form was obtained by
reversing the signs of both $U_N$ and $I_S$: $f_N(E,-U_N,-I_S) -
f_N(E,-U_N,I_S)$. The two data sets are in near perfect agreement.

If we increase $U_N$, we can convert the SNS JJ to a
$\pi$-junction \cite{Huang:02}.  The SNS JJ crosses over from the
0-state to the $\pi$-state at $U_N = 30~\mu$V.  The right panel of
Fig.~\ref{Fig4} shows $\delta f_N(E)$ data in the $\pi$-state for
$U_N = 53~\mu$V and $I_S = 0.9 I_c$. Compared to Fig.~4a, the sign
of the low-energy feature in $\delta f_N(E)$ is reversed,
demonstrating that the phase difference $\phi$, rather than the
supercurrent, determines the sign of the effective temperature
gradient across the SNS JJ.

Let us compare these results to those obtained from
Eqs.~\eqref{eq:kinetic} using the parameters $E_T$, $\Delta$
obtained from the previous fits, with no additional fitting. The
computed theory curves are shown as solid lines in
Fig.~\ref{Fig4}, and they agree with the experimental data rather
well. Note that for this calculation, we entirely neglected the
effect of inelastic scattering.

As stated above, we have recovered $\delta f_N(E)$ without knowing
the exact form of $n_N(E)$.  We have verified that our results are
rather insensitive to small changes in $n_N(E)$ by performing the
deconvolution using several different forms for $n_N(E)$, with the
value of $n_N(0)$ varying from 0.1 to 0.5.  The amplitude of the
oscillations in $\delta f_N(E)$ are inversely proportional to
$n_N(E)$ near $E=0$, but the overall shape does not change from
those shown in Fig.~\ref{Fig4}.  To extract the full $f_N(E)$ from
our data would require a separate measurement of $n_N(E)$ (using a
normal tunnel probe \cite{SophiePRL}) at each value of $U_N$ and
$I_S$ used in our experiment. As $U_N$ increases, deviations of
$n_N(E)$ from its equilibrium form may arise from nonequilibrium
processes in the superconducting reservoirs, where quasiparticle
current is converted to supercurrent \cite{Mercereau:72}.

The feature we have measured has been called a ``Peltier-like"
effect in Ref.~\onlinecite{Tero:03}, because an applied current
induces a temperature gradient. It should be emphasized, however,
that our experiment is far from equilibrium. We are not measuring
a linear response transport coefficient, as was done with the
thermopower of Andreev interferometers
\cite{Dikin:02,Thermopowerfootnote}. It is unlikely that one could
achieve real local cooling in our experiment, as one does with the
conventional Peltier effect. Perhaps the most intriguing aspect of
our results is that the nonequilibrium feature in $f(E)$ shown in
Fig.~4 provides an approximate visualization of the spectral
supercurrent density $j_E$ -- a quantity whose energy dependence
has previously been inferred only indirectly through
$\pi$-junction experiments.

We are grateful to F. Wilhelm, F. Pierre, and S. Yip for fruitful
discussions. N.B. also thanks H. Pothier and D. Esteve for their
long-standing collaboration. This work was supported by NSF grants
DMR-0104178 and 0405238, by the Keck Microfabrication Facility
supported by NSF DMR-9809688, and by the Academy of Finland.

\end{document}